\documentclass[11pt]{article}

\usepackage[T1]{fontenc}
\usepackage{amsmath}
\usepackage{booktabs}
\usepackage{graphicx}
\usepackage{longtable}
\usepackage{multirow}
\usepackage[numbers,sort&compress]{natbib}
\usepackage{subcaption}
\usepackage[margin=1in]{geometry}
\usepackage{microtype}
\usepackage[hidelinks,hypertexnames=false]{hyperref}

\title{Standardized Planck-Mean Absorption Coefficients for High-Temperature Gases}
\author{Ryan D. DeBoskey\textsuperscript{1,*},
Clayton M. Geipel\textsuperscript{2}, and
Brian T. Bojko\textsuperscript{1}\\[0.6em]
\small\textsuperscript{1}Laboratories for Computational Physics and Fluid Dynamics,\\[-0.1em]
\small U.S. Naval Research Laboratory, Washington, DC 20375, USA\\[0.3em]
\small\textsuperscript{2}Combustion and Reacting Transport Section,\\[-0.1em]
\small U.S. Naval Research Laboratory, Washington, DC 20375, USA\\[0.5em]
\small\textsuperscript{*}\href{mailto:ryan.d.deboskey.ctr@us.nrl.navy.mil}{ryan.d.deboskey.ctr@us.nrl.navy.mil}}
\date{}

\begin{document}
\maketitle

\begin{abstract}
Planck-mean absorption coefficients of pure gases are calculated from line-by-line integrations of HITEMP and ExoMol spectra up to 5000 K for use in high-enthalpy radiative heat transfer simulations.  The dataset extends beyond conventional combustion products, including high-temperature radicals, nitrogen-bearing species, soot precursors, metal oxides, metal hydrides, and halides: AlF, AlH, AlO, BH, C$_2$, C$_2$H$_2$, CH, CH$_4$, CN, CO, CO$_2$, H$_2$O, HBO, HCN, HF, MgH, MgO, N$_2$O, NO, NO$_2$, OH, SiO, TiH, TiO.
New calculations are validated against historical correlations for overlapping species, demonstrating good agreement over their applicable temperature ranges.  Strong visible opacities result in large Planck-mean absorption coefficients for many of the considered species, highlighting the need to account for their radiative contribution in extreme environments.
A brute-force search over several candidate functional forms is performed to provide a common set of coefficients, establishing a standardized framework that can be easily integrated into production solvers and extended as additional high-temperature line lists become available.


\end{abstract}


\section{Introduction}

Radiative heat transfer is often the dominant mode of energy transport in high-enthalpy and extreme thermal environments, yet its accurate prediction is challenged by the immense spectral complexity of high-temperature gases \cite{modest2021radiative}.  While traditional analyses focus on standard combustion products, modern aerospace and industrial applications demand the inclusion of transient radicals, soot precursors, and metal oxides whose massive, dense line lists make rigorous line-by-line (LBL) calculations computationally prohibitive \cite{ROTHMAN20102139}.
Solving the radiative transfer equation (RTE) in large unsteady computational fluid dynamics (CFD) simulations thus requires efficient spectral models, with most engineering codes relying on reduced gray-gas properties (most notably, Planck-mean absorption coefficients) \cite{FUREBY20213021,GUHATHAKURTA2021104509}.

Legacy Planck-mean absorption coefficient correlations for combustion gases trace largely to older narrow-band or semi-empirical workflows such as those summarized by Tien, Barlow et al., and the RADCAL framework \cite{tien1968thermal,BARLOW20012102,grosshandler1993radcal}. Those correlations remain immensely valuable and enable radiation modeling to be widely accessible and rapidly integrated into engineering solvers.  However, these correlations are tied to the spectroscopic coverage and fitting assumptions available at the time of creation, a limited number of species (CO, CO$_2$, H$_2$O, CH$_4$), and narrow temperature ranges (largely under 2500 K). Subsequent re-evaluations using higher-resolution spectroscopic databases and LBL calculations have shown meaningful differences relative to older tabulations \cite{ZHANG2002649,RIVIERE20123349,chu2014calculations,chmielewski2015planck,zheng2022co2}.  This represents a critical bottleneck hindering widespread and accurate RTE calculations, as CFD practitioners are currently required to search through the literature for individual, idiosyncratic species correlations, perform their own ad-hoc LBL calculations \cite{KANEKO2026105997}, or neglect important radiative species altogether.

For major combustion products, HITEMP provides the standard high-temperature foundation, rigorously resolving the hot-band transitions required for flames, exhaust, and other thermally excited gas mixtures (CH$_4$, CO$_2$, CO, H$_2$O, N$_2$O, NO$_2$, NO, and OH) \cite{ROTHMAN20102139,Hargreaves2019SpectroscopicLine}. ExoMol, a modern line list database initially created for studying exoplanets, extends this capability by providing spectral information to a wider array of species that are sparse or entirely absent in traditional combustion databases, encompassing radicals, metal oxides, hydrides, halides, and soot precursors \cite{MELIN2020107079,tennyson2012exomol}.
Expanding this species set is critical for the current generation of propulsion and hypersonic applications, which require modeling advanced fuels, energetic additives, and ablative materials.
Few existing Planck-mean absorption coefficient correlations are even valid at the high temperatures observed within propulsion and hypersonic systems \cite{RIVIERE20123349}.
Furthermore, with increasing temperature, the Planck function weighting shifts toward shorter wavelengths, meaning other species beyond the conventional CO$_2$ and H$_2$O pair can become radiatively dominant, notably metal oxides and hydrides \cite{Gandhi2019}.


The present work aims to provide a standardized set of Planck-mean absorption coefficient correlations for easy integration within production-ready CFD solvers.  The objective of this paper is not only to generate new Planck-mean absorption coefficient data, but to rigorously reduce and standardize the data into compact functional forms suitable for practical implementation.  The use of unified reduced-order property representations has a highly successful precedent in democratizing modeling (such as in species chemical thermodynamics \cite{Mcb93,Mcb02}), and the RTE modeling community stands to benefit from a similar framework.

The remainder of this paper is organized as follows.  Section \ref{methodology_section} details the spectroscopic database coverage, LBL calculation procedure, and brute-force search methodology for the best reduced-order representation.  The results are presented and discussed in Section \ref{results_section}, showcasing the new LBL calculations and final tabulated coefficients for all 24 species.  Finally, conclusions are drawn in Section \ref{conclusion_section}.

\section{Methodology}\label{methodology_section}

\subsection{Database coverage}


For this work, all line lists for relevant gaseous species are gathered from the HITEMP \cite{ROTHMAN20102139} and ExoMol databases \cite{tennyson2012exomol}.  Species within the HITRAN database are purposefully not considered \cite{GORDON2022107949}, as the goal of this work is to provide Planck-mean absorption coefficient fits for use in high-temperature applications.  Species are selected based on three criteria:

\begin{enumerate}
  \item Line lists are valid over application-relevant high-temperature ranges, generally extending to 5000 K.
  \item Line lists contain a large enough (or complete) spectral range such that the Planck-mean absorption coefficient remains valid at higher temperatures (blackbody weighting shift toward higher wavenumbers).
  \item Species are relevant for combustion, propulsion, hypersonics, and/or industrial CFD applications.
\end{enumerate}

Criterion 1 is a practical limit in high-temperature configurations, as the total radiative emission scales with temperature to the fourth power ($T^4$), and the radiative contribution of low-temperature species is dwarfed by the emissions of the high-temperature product and radical species.  While past efforts have compiled Planck-mean absorption coefficients for parent hydrocarbon fuels at lower temperatures \cite{Wakatsuki18022008}, such complex molecular species are intentionally excluded from the present database.  At the elevated temperatures of interest in this study (up to 5000 K), heavy parent hydrocarbons are highly unstable and rapidly pyrolyze into smaller radical fragments and lighter product species.
Consequently, by omitting such transient, low-temperature species and focusing on the high-temperature products and abundant radicals, we maintain a highly focused, computationally efficient database without compromising the fidelity of RTE calculations.


\begin{table}[htbp]
\centering
\caption{Summary of spectroscopic databases used for calculating Planck-mean absorption coefficients.}
\label{tab:current_species}
\footnotesize
\begin{tabular}{c c l l c l}
\toprule
Species & Database & $T_{\max}$ [K] & $N_{\mathrm{lines}}^{*}$ & $\frac{\nu_{\max}}{1000}$ [cm$^{-1}$] & Reference \\
\midrule
\(\mathrm{AlF}\) & ExoMol & 5000 & 40.4 k & 3.88 & \cite{BERNATH2020106687} \\
\(\mathrm{AlH}\) & ExoMol & 5000 & 29.7 k & 24.6 & \cite{AlH_2024} \\
\(\mathrm{AlO}\) & ExoMol & 5000 & 4.94 M & 35.0 & \cite{AlO_exomol_1,AlO_exomol_2} \\
\(\mathrm{BH}\) & ExoMol & 5000 & 5.53 k & 28.5 & \cite{tennyson2026bh} \\
\(\mathrm{C_2}\) & ExoMol & 5000 & 6.08 M & 40.0 & \cite{exomol_C2_OG,exomol_C2} \\
\(\mathrm{C_2H_2}\) & ExoMol & 2500 & 4.34 B & 10.0 & \cite{exomol_acetylene} \\
\(\mathrm{CH}\) & ExoMol & 5000 & 53.1 k & 39.2 & \cite{masseron2014ch,BERNATH2020106687} \\
\(\mathrm{CH_4}\) & HITEMP & 2500 & 31.8 M & 13.4 & \cite{Hargreaves2020MethaneLine} \\
\(\mathrm{CN}\) & ExoMol & 5000 & 5.00 M & 60.0 & \cite{kozlov2024upgraded_CN} \\
\(\mathrm{CO}\) & HITEMP & 5000 & 752 k & 22.2 & \cite{Li2015RovibrationalLine} \\
\(\mathrm{CO_2}\) & HITEMP & 5000 & 326 M & 17.7 & \cite{Hargreaves2025CarbonDioxide} \\
\(\mathrm{H_2O}\) & HITEMP & 5000 & 114 M & 30.0 & \cite{ROTHMAN20102139} \\
\(\mathrm{HBO}\) & ExoMol & 5000 & 75.3 M & 9.00 & \cite{HBO_exomol} \\
\(\mathrm{HCN}\) & ExoMol & 4000 & 34.4 M & 17.5 & \cite{exomol_HCN_OG,exomol_HCN} \\
\(\mathrm{HF}\) & ExoMol & 5000 & 7.96 k & 32.3 & \cite{COXON2015133,Somogyi2021} \\
\(\mathrm{MgH}\) & ExoMol & 5000 & 88.5 k & 29.6 & \cite{MgH_exomol} \\
\(\mathrm{MgO}\) & ExoMol & 5000 & 72.8 M & 37.0 & \cite{exomol_MgO} \\
\(\mathrm{N_2O}\) & HITEMP & 5000 & 3.63 M & 12.9 & \cite{Hargreaves2019SpectroscopicLine} \\
\(\mathrm{NO}\) & HITEMP & 5000 & 1.14 M & 26.8 & \cite{Hargreaves2019SpectroscopicLine} \\
\(\mathrm{NO_2}\) & HITEMP & 5000 & 1.11 M & 4.78 & \cite{Hargreaves2019SpectroscopicLine} \\
\(\mathrm{OH}\) & HITEMP & 5000 & 57.0 k & 43.4 & \cite{GORDON2022107949} \\
\(\mathrm{SiO}\) & ExoMol & 5000 & 91.3 M & 72.0 & \cite{SiO_lines} \\
\(\mathrm{TiH}\) & ExoMol & 5000 & 199 k & 24.0 & \cite{burrows2005spectroscopic,BERNATH2020106687} \\
\(\mathrm{TiO}\) & ExoMol & 5000 & 59.0 M & 30.0 & \cite{hybridTiO,TiO_exomol} \\
\bottomrule
\multicolumn{6}{l}{\scriptsize $^{*}$ k = thousand, M = million, B = billion.} \\
\end{tabular}
\end{table}

Table~\ref{tab:current_species} summarizes the species considered in this study, along with their source database.  For all species, the reported line counts are taken directly from the downloaded line-list metadata.
As emphasized, the database extends beyond conventional combustion products, including high-temperature radicals, nitrogen-bearing species, soot precursors, metal oxides, metal hydrides, and halides: AlF, AlH, AlO, BH, C$_2$, C$_2$H$_2$, CH, CH$_4$, CN, CO, CO$_2$, H$_2$O, HBO, HCN, HF, MgH, MgO, N$_2$O, NO, NO$_2$, OH, SiO, TiH, TiO.

Most species are calculated up to 5000 K. HCN is available through 4000 K, while methane (CH$_4$) and acetylene (C$_2$H$_2$) are calculated up to 2500 K (Criteria 1 and 2).  By a similar physical rationale as above, these lighter hydrocarbon species are significantly more thermally robust than larger fuels and serve as important intermediates during the thermal dissociation and oxidation of large hydrocarbons \cite{Wakatsuki18022008}.  They themselves undergo rapid thermal dissociation and oxidation at high temperatures, limiting their meaningful physical presence to regimes below 2500 K.

\subsection{Planck-mean absorption coefficient calculation}

The Planck-mean absorption coefficient, $a_{\mathrm p}(T)$, is evaluated line-by-line \cite{ZHANG2002649,RIVIERE20123349,chu2014calculations,chmielewski2015planck,zheng2022co2}:

\begin{equation}
a_{\mathrm p}(T) =
\frac{\int_{\nu_{\min}}^{\nu_{\max}} \kappa_{\nu}(T) I_{\mathrm b}(\nu, T) \,\mathrm{d}\nu}
{\int_{0}^{\infty} I_{\mathrm b}(\nu, T) \,\mathrm{d}\nu}
\label{eq:planck_mean}
\end{equation}

\noindent where $\nu$ is the wavenumber, $\kappa_{\nu}$ is the monochromatic absorption coefficient, $T$ is the temperature, and $I_{\mathrm b}(\nu, T)$ is the blackbody spectral radiance.  The blackbody weighting function is evaluated from Planck's law:

\begin{equation}
I_{\mathrm b}(\nu, T) =
\frac{2 h c^2 \nu^3}{\exp \left(\dfrac{h c \nu}{k_{\mathrm B} T}\right)-1}
\label{eq:planck_weight}
\end{equation}

\noindent where $h$ is Planck's constant, $c$ is the speed of light in a vacuum, and $k_{\mathrm B}$ is the Boltzmann constant.  Equation \ref{eq:planck_mean} is then simplified to:

\begin{equation}
a_{\mathrm p}(T) = \frac{\pi}{\sigma T^4} \int_{\nu_{\min}}^{\nu_{\max}} \kappa_{\nu}(T) I_{\mathrm b}(\nu, T) \,\mathrm{d}\nu
\label{eq:planck_mean2}
\end{equation}

\noindent where $\sigma$ is the Stefan-Boltzmann constant.  The spectral absorption coefficient, $\kappa_{\nu}$, at a specific wavenumber $\nu$ is calculated by summing the influence of all individual spectral lines \cite{zheng2022co2}:

\begin{equation}
\kappa_{\nu} = N \sum_{i} S_{i}(T) F_{i}(\nu)
\label{eq:absorption_coeff}
\end{equation}

\noindent where $N$ is the molecular number density of the absorbing gas, $S_{i}(T)$ is the line intensity of the $i^{\mathrm{th}}$ transition at temperature $T$, and $F_{i}(\nu)$ is the line shape profile function for the $i^{\mathrm{th}}$ transition.  The line intensity $S_{i}(T)$ quantifies the strength of a given absorption line and is highly dependent on temperature. It can be calculated from a known value at a reference temperature, $T_{\mathrm{ref}}$ \cite{RIVIERE20123349}:

\begin{equation}
S_{i}(T) = S_{i}(T_{\mathrm{ref}}) \frac{Q(T_{\mathrm{ref}})}{Q(T)} \exp\left[ \frac{h c E''}{k_{\mathrm B}} \left( \frac{1}{T_{\mathrm{ref}}} - \frac{1}{T} \right) \right] \frac{1 - \exp\left(-\frac{h c \nu_{0,i}}{k_{\mathrm B} T}\right)}{1 - \exp\left(-\frac{h c \nu_{0,i}}{k_{\mathrm B} T_{\mathrm{ref}}}\right)}
\label{eq:line_intensity}
\end{equation}

\noindent where $S_{i}(T_{\mathrm{ref}})$ is the line intensity at the reference temperature $T_{\mathrm{ref}}$, $Q(T)$ is the total internal partition function of the molecule at temperature $T$, $E''$ is the lower-state energy of the transition, and $\nu_{0,i}$ is the transition line-center wavenumber.  To ensure methodological consistency, the partition function for all species was evaluated via a full direct summation over the complete set of energy levels provided within each respective database.

The spectral integrations are performed numerically using Simpson's rule with a constant spectral step of $\Delta \nu = 0.01$ cm$^{-1}$ \cite{RIVIERE20123349,zheng2022co2}. Transitions with temperature-dependent line intensities below $S_{\mathrm{cut}}=10^{-30}$ cm molecule$^{-1}$ were omitted.
A Voigt line-shape profile, $F_{i}(\nu)$, is applied with a line-wing cutoff of $\Delta \nu_{\mathrm{cut}}$ = 100 cm$^{-1}$ \cite{RIVIERE20123349}.
All calculations are performed at a total pressure of 1 atm in 50 K increments over each species-specific temperature range described above ($a_{\mathrm p}$ reported in pressure-normalized units m$^{-1}$ atm$^{-1}$).
Because the integral of the Voigt line shape function is normalized to unity, the global spectral integral defining the Planck-mean absorption coefficient is largely insensitive ($\le$ 1\%) to pressure-induced collisional broadening strictly in the optically thin limit (low to moderate pressures) \cite{modest2021radiative}.
The monochromatic spectra are evaluated over the full spectral range spanned by each species ($\nu_{\min} \le \nu \le \nu_{\max}$), with a $\nu_{\max}$ listed in Table \ref{tab:current_species} per species and $\nu_{\min}$ = 1 cm$^{-1}$.
This dynamic range ensures extended resolution far into the visible and near-ultraviolet regions to capture the electronic transition bands of the considered refractory metal oxides and radicals.
For computational efficiency, only the primary isotope of each species is calculated and the resulting spectral absorption is scaled to an isotopic abundance of unity (100\%).
The adopted line-strength cutoff, line-wing extent, spectral spacing, pressure, and line list treatment were assessed and selected based on sensitivity calculations performed on CO summarized in Appendix~\ref{app:sensitivity}.
Newly calculated fits are compared to legacy fits \cite{BARLOW20012102,grosshandler1993radcal}, high-temperature fits \cite{RIVIERE20123349}, and contemporary fits \cite{zheng2022co2}.

\subsection{Selection of reduced-order representation}

To establish a standardized and computationally efficient representation, a brute-force optimization search was conducted over four candidate functional families. These forms relate the tabulated Planck-mean absorption coefficients, $a_{\mathrm p}$, to the local gas temperature, $T$ [K]. In all formulations, the natural logarithm ($\ln$) and natural exponential ($\exp$) are utilized.  The candidate functional families are defined as follows:

\begin{equation}\tag{F1}
    a_{\mathrm p}=\sum_n c_n T^n
\end{equation}

\begin{equation}\tag{F2}
    a_{\mathrm p}=\sum_n c_n (1000/T)^n
\end{equation}

\begin{equation}\tag{F3}
    a_{\mathrm p}=\exp\left[\sum_n c_n (1000/T)^n\right]
\end{equation}

\begin{equation}\tag{F4}
    a_{\mathrm p}=\exp\left[\sum_n c_n \left(\ln(T/1000\,\mathrm{K})\right)^n\right]
\end{equation}

\noindent where $n$ is the degree of the polynomial fit, and $c_n$ are the corresponding coefficients for the $n^{\mathrm{th}}$-degree terms.  The brute-force search is performed across all four candidate functions with polynomial degrees $n \in [3,6]$.  Each candidate function is fit using a single ``1-zone'' and ``2-zone'' temperature break fit, where the 2-zone temperature break is determined from $T_{\mathrm{break}} \in [1000, 1500, 2000, 2500]$ K. The two polynomial branches are constrained to have the same value at the breakpoint ($C^0$ continuity).  The 2-zone break enables higher accuracy across the wide temperature range and has precedent in historical fits \cite{BARLOW20012102,grosshandler1993radcal,Mcb93,Mcb02}. Because the CH$_4$ and C$_2$H$_2$ datasets terminate at 2500 K, these two species retain single-zone fits. Each candidate function is evaluated based on mean and maximum root mean square (RMS) error.  The overall best RMS candidate function is not necessarily selected, as the relative reduction in error is considered with increasing polynomial order for maintaining model parsimony.




\section{Results}\label{results_section}

\subsection{Validation}

\begin{figure}[htbp]
    \centering
    \begin{subfigure}[t]{0.49\textwidth}
        \centering
        \includegraphics[width=\linewidth]{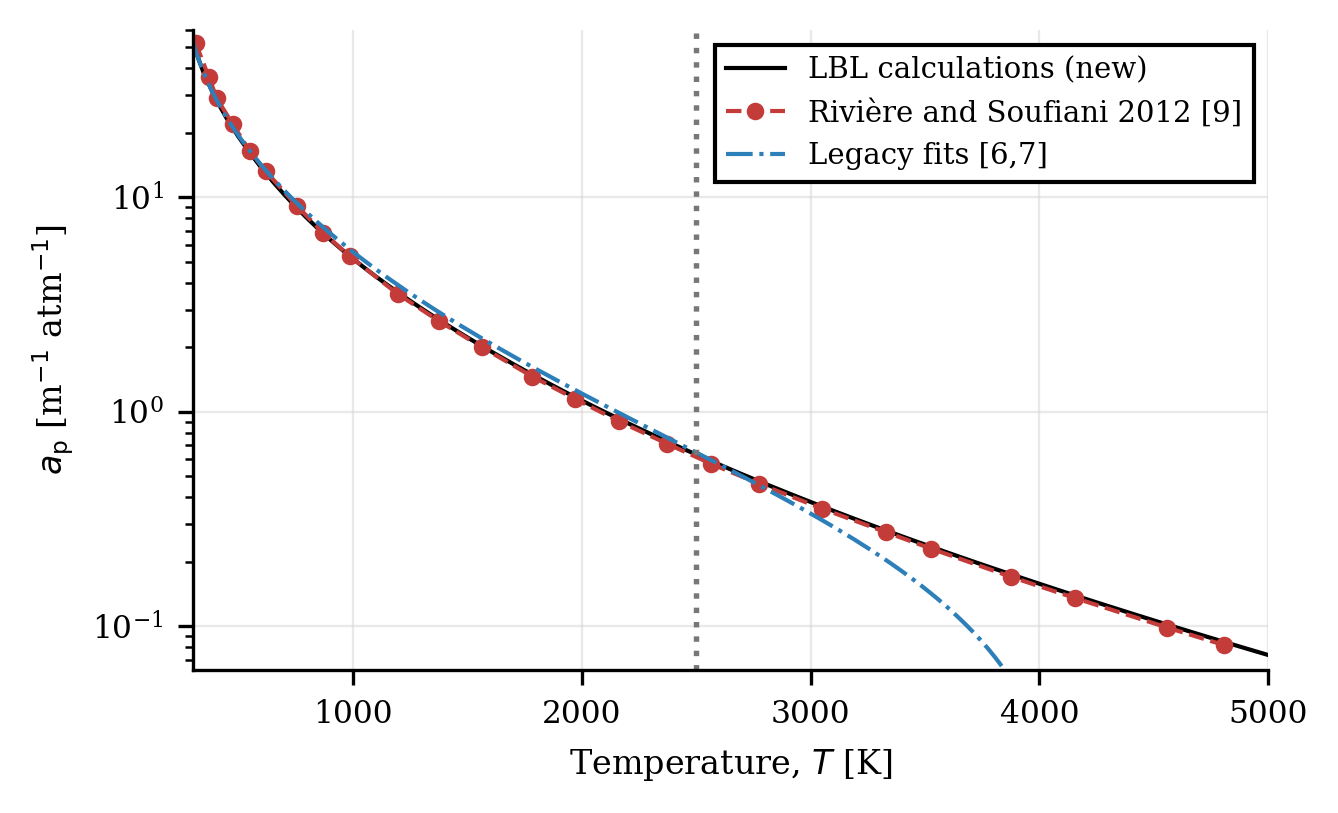}
        \caption{\(\mathrm{H_2O}\)}
    \end{subfigure}
    \hfill
    \begin{subfigure}[t]{0.49\textwidth}
        \centering
        \includegraphics[width=\linewidth]{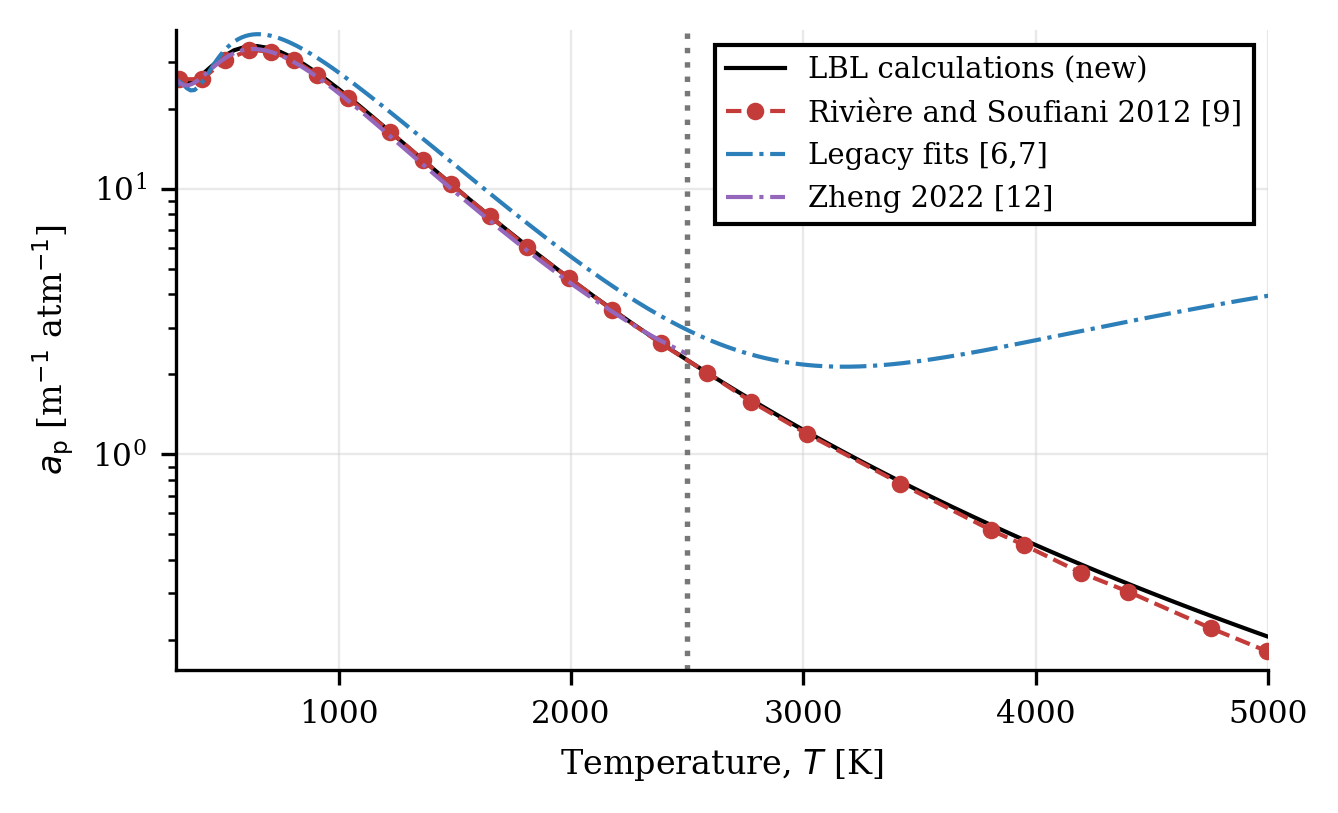}
        \caption{\(\mathrm{CO_2}\)}
    \end{subfigure}
    \hfill
    \begin{subfigure}[t]{0.49\textwidth}
        \centering
        \includegraphics[width=\linewidth]{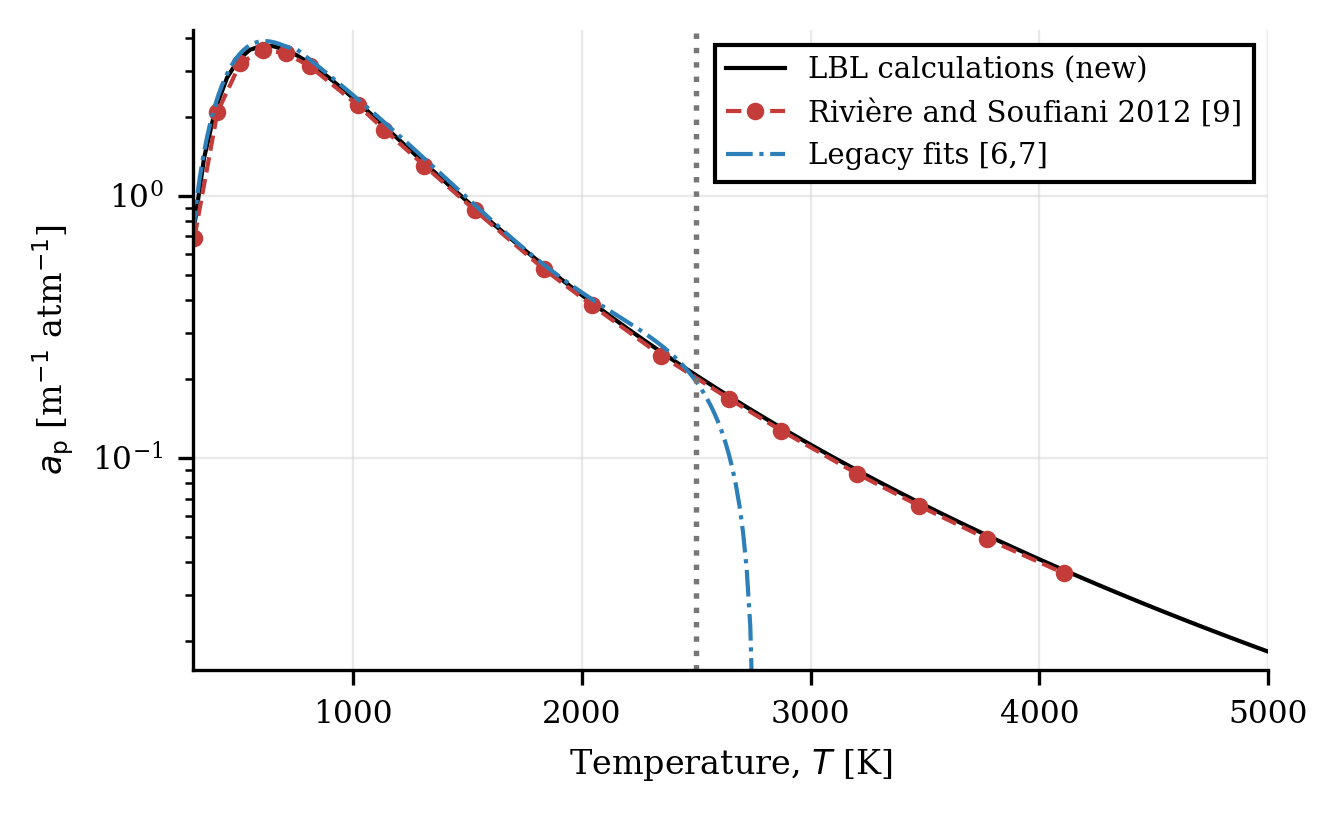}
        \caption{\(\mathrm{CO}\)}
    \end{subfigure}
    \begin{subfigure}[t]{0.49\textwidth}
        \centering
        \includegraphics[width=\linewidth]{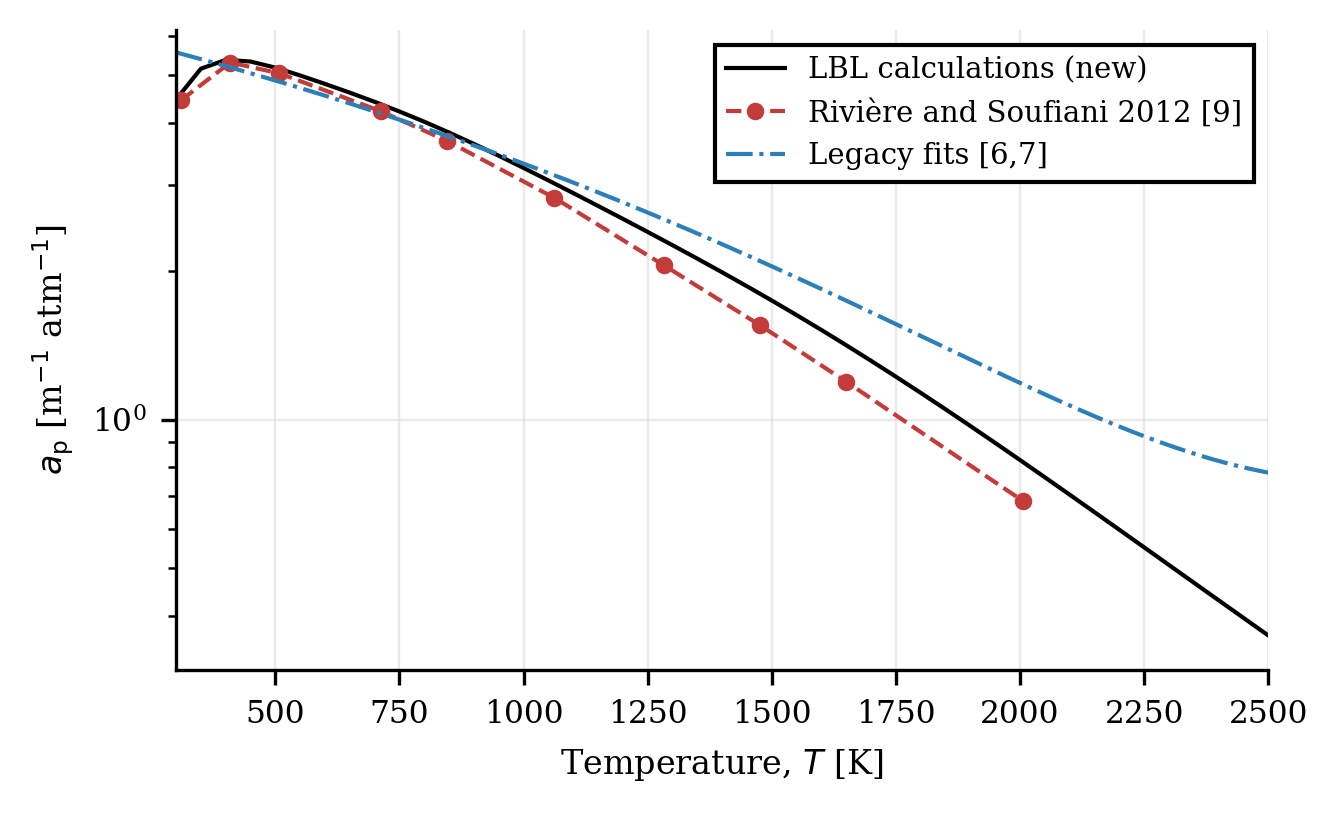}
        \caption{\(\mathrm{CH_4}\)}
    \end{subfigure}
    \caption{Validation comparison of newly calculated Planck-mean absorption coefficients with existing fits found in the literature.}
    \label{fig:validation_suite}
\end{figure}

Figure~\ref{fig:validation_suite} shows a comparison of the newly calculated Planck-mean absorption coefficients for H$_2$O, CO$_2$, CO, and CH$_4$ with existing fits found within the literature.  The agreement is very good over the overlapping temperature ranges, with the legacy fits showing a clear breakdown at 2500 K \cite{BARLOW20012102,grosshandler1993radcal} (the dotted vertical marker indicates the maximum valid fit temperature).
This limitation highlights the necessity of the current methodology, as polynomial approximations fundamentally fail to maintain physical fidelity when extrapolated beyond their calibration window.
The present line-by-line calculations extend the usable range to 5000 K (excluding CH$_4$), in accordance with high-temperature fits provided by Rivière and Soufiani \cite{RIVIERE20123349}.
Deviations are observed with the legacy fits \cite{BARLOW20012102,grosshandler1993radcal} owing to new LBL calculation methodologies, and minor deviations with Rivière and Soufiani \cite{RIVIERE20123349} owing to updated database coverage since 2012.  The present CO$_2$ calculations closely align with contemporary fits from Zheng et al. (2022) \cite{zheng2022co2}, validating the current methodology.

\subsection{New LBL calculations}

\begin{figure}[htbp]
    \centering
    \begin{subfigure}[t]{0.48\textwidth}
        \centering
        \includegraphics[width=\linewidth]{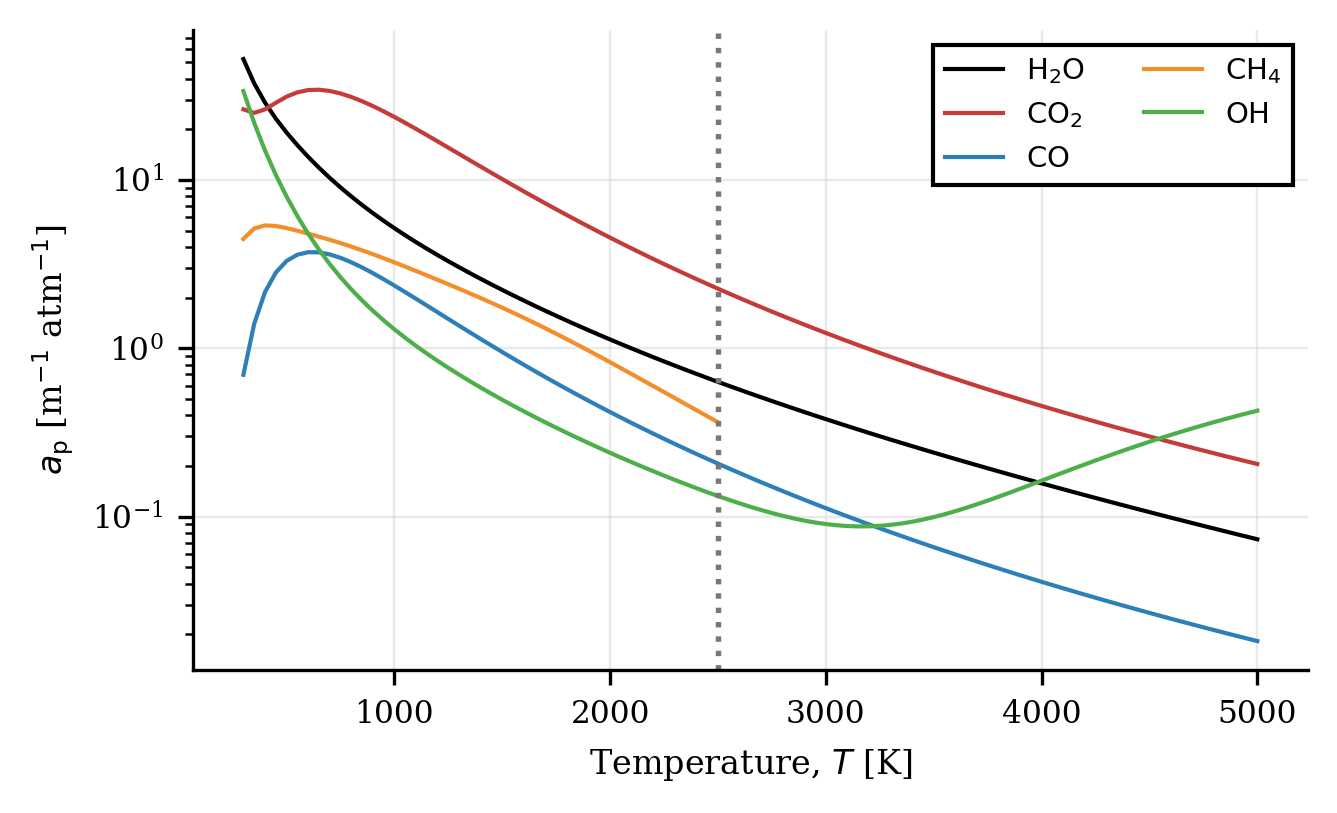}
        \caption{Major combustion products}
    \end{subfigure}
    \hfill
    \begin{subfigure}[t]{0.48\textwidth}
        \centering
        \includegraphics[width=\linewidth]{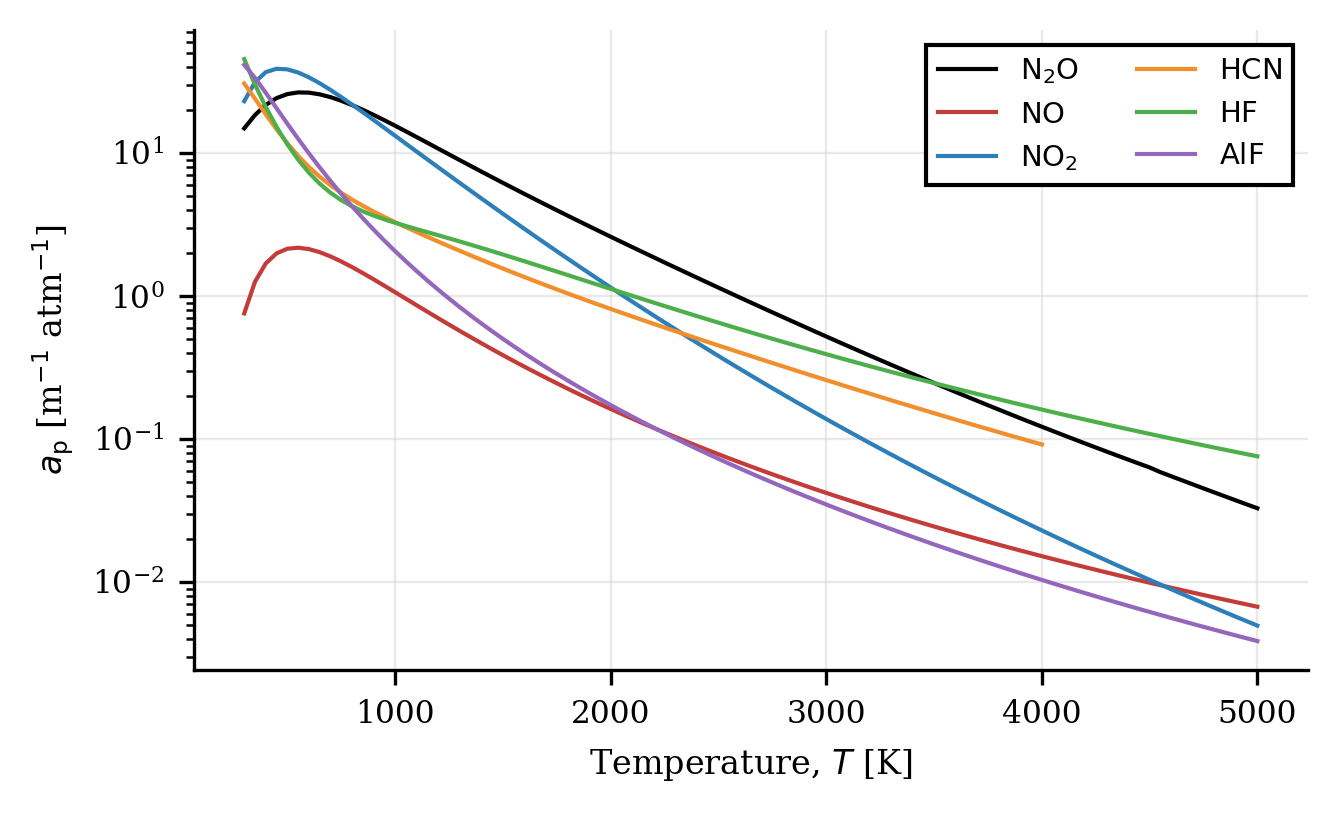}
        \caption{Nitrogen-bearing species and halides}
    \end{subfigure}

    \medskip

    \begin{subfigure}[t]{0.48\textwidth}
        \centering
        \includegraphics[width=\linewidth]{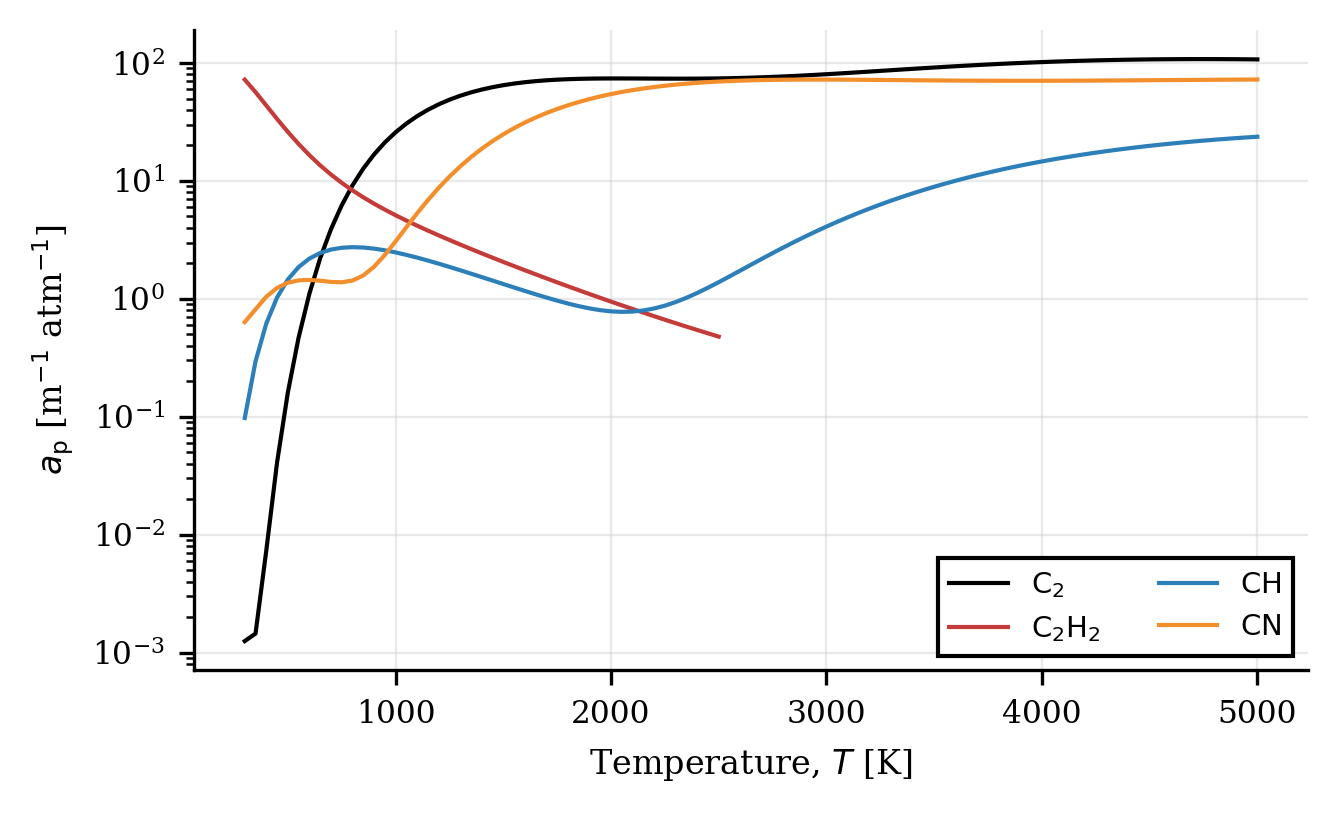}
        \caption{Soot precursors and carbon radicals}
    \end{subfigure}
    \hfill
    \begin{subfigure}[t]{0.48\textwidth}
        \centering
        \includegraphics[width=\linewidth]{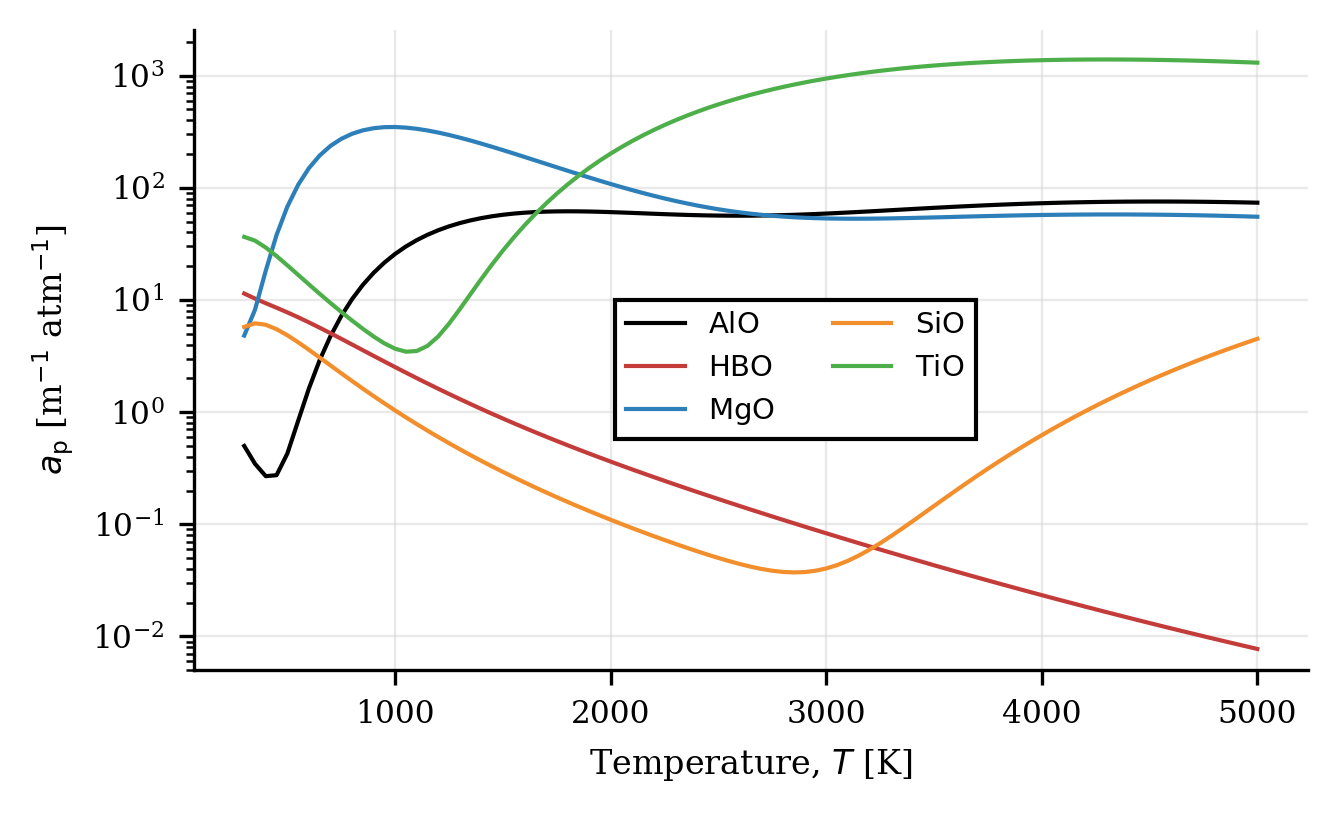}
        \caption{Refractory oxides}
    \end{subfigure}
    \medskip
    \begin{subfigure}[t]{0.48\textwidth}
        \centering
        \includegraphics[width=\linewidth]{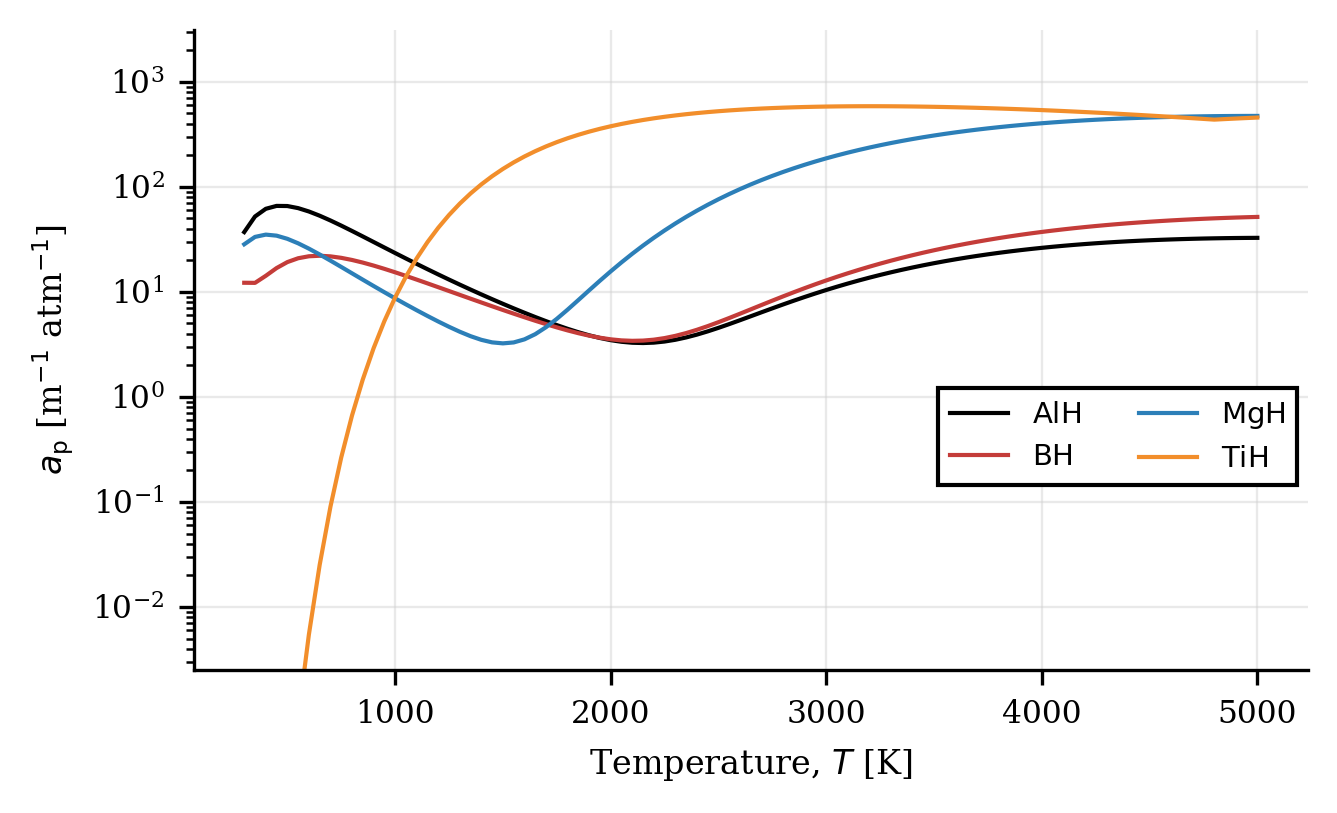}
        \caption{Refractory hydrides}
    \end{subfigure}
    \caption{Grouped line-by-line Planck-mean absorption coefficients for the expanded species library. The grouped presentation emphasizes how the dominant absorption strength shifts with temperature toward nontraditional emitters.}
    \label{fig:lbl_grouped}
\end{figure}

Figure~\ref{fig:lbl_grouped} shows the grouped line-by-line Planck-mean absorption coefficients for the expanded species library.  Overall, the computed trends reveal a shift in dominance from stable polyatomic molecules to diatomic radicals and metal-bearing species with increasing temperatures.  For stable polyatomic molecules, absorption is dominated by infrared rovibrational transitions.  Owing to Wien's Law ($\nu_{\max} \propto T$), at lower temperatures the blackbody spectrum peaks in the far-infrared ($\approx 1000\,\mathrm{cm}^{-1}$), matching the strong vibrational-rotational bands of stable molecules. At temperatures of $3000\text{--}5000\,\mathrm{K}$, the blackbody spectrum shifts directly into the near-infrared, visible, and near-ultraviolet regions ($10000\text{--}25000\,\mathrm{cm}^{-1}$).  Stable polyatomics do not possess strong electronic transition bands in the visible or near-ultraviolet regions, and the shifting Planck weighting function moves away from their active bands, resulting in lower $a_{\mathrm p}(T)$ values. Conversely, diatomic radicals ($\text{C}_2$, $\text{CN}$) and transition-metal oxides ($\text{TiO}$, $\text{AlO}$) contain unpaired outer-shell electrons that give rise to exceptionally strong, highly dense electronic transition band systems in the visible and near-infrared, such as the Swan bands of $\text{C}_2$, the Violet and Red systems of $\text{CN}$, and the massive $\alpha$, $\beta$, and $\gamma$ systems of $\text{TiO}$.

The major combustion group comprises the primary reactants, intermediates, and stable products of hydrocarbon-air flames, specifically $\text{H}_2\text{O}$, $\text{CO}_2$, $\text{CO}$, $\text{CH}_4$, and $\text{OH}$.  Water vapor and carbon dioxide are the major absorbers within this group.  Owing to its low equilibrium mole fraction at low flame temperatures, $\text{OH}$ has been typically neglected in mixture radiative transport models.
This group of gaseous species, along with soot, dominate within traditional combustion systems operating at moderate temperatures and have been the focus of past RTE modeling \cite{ZHANG2002649,RIVIERE20123349,chu2014calculations,chmielewski2015planck,zheng2022co2}.

The nitrogen-bearing and halogenated group includes $\text{N}_2\text{O}$, $\text{NO}$, $\text{NO}_2$, $\text{HCN}$, $\text{HF}$, and $\text{AlF}$. Within this group, nitrogen dioxide ($\text{NO}_2$) dominates the low-temperature regime, reaching its peak near $500\,\mathrm{K}$ before undergoing rapid decay.  Hydrogen fluoride ($\text{HF}$) starts with a high absorption coefficient of $\approx 50\,\mathrm{m}^{-1}\,\mathrm{atm}^{-1}$ at $300\,\mathrm{K}$ and decays less rapidly due to a large rotational constant and high fundamental frequency \cite{HF_large_ap,10.1115/1.1416689}, becoming the dominant absorber in this group above 2500 K and a critical radiative species in hydrofluorocarbon flames \cite{KANEKO2026105997}.  The soot precursors and carbon radicals feature $\text{C}_2$, $\text{C}_2\text{H}_2$, $\text{CH}$, and $\text{CN}$. Acetylene ($\text{C}_2\text{H}_2$), as a stable, closed-shell polyatomic molecule with high infrared activity, dominates the low-temperature regime. The radicals $\text{C}_2$, $\text{CH}$, and $\text{CN}$ show an increase of several orders of magnitude in Planck-mean absorption coefficient with increasing temperature, reaching much higher values than the typically considered major combustion products at high temperatures.
This group of species is representative of the critical radiative contributors for hypersonic reentry as the surrounding air dissociates and the thermal protection system (TPS) undergoes ablation \cite{salas_hypersonics_review}.


The refractory metal-bearing group, comprising $\text{AlH}$, $\text{AlO}$, $\text{BH}$, $\text{HBO}$, $\text{MgH}$, $\text{MgO}$, $\text{SiO}$, $\text{TiH}$, and $\text{TiO}$, is characterized by absorption magnitudes that are orders of magnitude higher than standard combustion gases.  In contrast to weaker vibrational-rotational transitions within the ground electronic states of $\text{CO}_2$ and $\text{H}_2\text{O}$, the refractory species contain strong electronic transition bands in the visible or near-ultraviolet regions.
High-speed propulsion systems frequently incorporate metals such as aluminum, boron, or magnesium as propellant or fuel additives to increase specific impulse.  Accurate Planck-mean data for these metal-bearing species are necessary for predicting radiative heating within combustors, nozzles, and plumes.

It is important to note the implications of the large absorption magnitudes for practical engineering simulations.  The Planck-mean absorption coefficients reported here are fundamental gas properties integrated over the optically thin limit.  Mixtures containing even trace mole fractions of species with large cross-sections, such as TiO, AlO, and C$_2$, will rapidly become optically thick ($\tau \ge 1$) over short geometric path lengths.  Consequently, practitioners must avoid applying these coefficients in simple optically-thin models which lack self-absorption corrections and will unphysically overpredict emission. The tabulated $a_{\mathrm p}$ coefficients should instead be utilized as the gray-gas absorption coefficients within RTE solvers that account for optical thickness and spatial self-absorption.


\subsection{Best-fit coefficients}

\begin{table}[htbp]
\centering
\caption{Summary of top performing functional fits}
\label{tab:fit_error_summary}
\footnotesize
\begin{tabular}{c c c c c c}
\toprule
Function & $n$ & Zones & $T_{\mathrm{break}}$ [K] & $\overline{\mathrm{RMS}}$ &  $\mathrm{RMS}_{\mathrm{max}}$  \\
\midrule
F3 & 6 & 2 & 1500 & 0.0096 & 0.0506 \\
F4 & 6 & 2 & 1500 & 0.0090 & 0.0573 \\
F4 & 5 & 2 & 2000 & 0.0170 & 0.0727 \\
F4\rlap{$^{*}$} & 5 & 2 & 1500 & 0.0145 & 0.0815 \\
F3 & 5 & 2 & 1500 & 0.0201 & 0.1024 \\
F4 & 4 & 2 & 1500 & 0.0300 & 0.1219 \\
\bottomrule
\multicolumn{6}{l}{\scriptsize $^{*}$Selected final representation.} \\
\end{tabular}
\end{table}

\begin{figure}[htbp]
    \centering
    \begin{subfigure}[t]{0.48\textwidth}
        \centering
        \includegraphics[width=\linewidth]{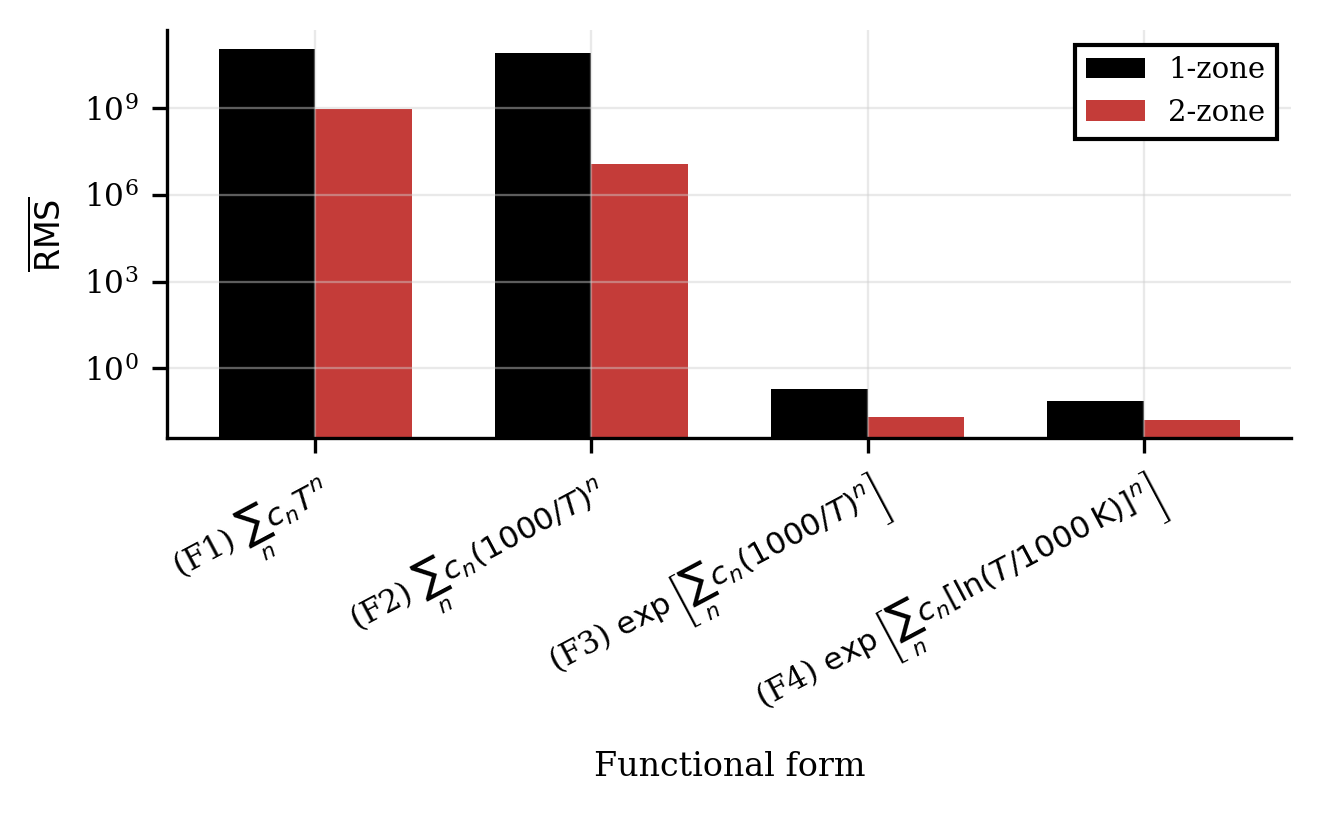}
        \caption{Mean fit error by functional form ($n$ = 5)}
    \end{subfigure}
    \hfill
    \begin{subfigure}[t]{0.48\textwidth}
        \centering
        \includegraphics[width=\linewidth]{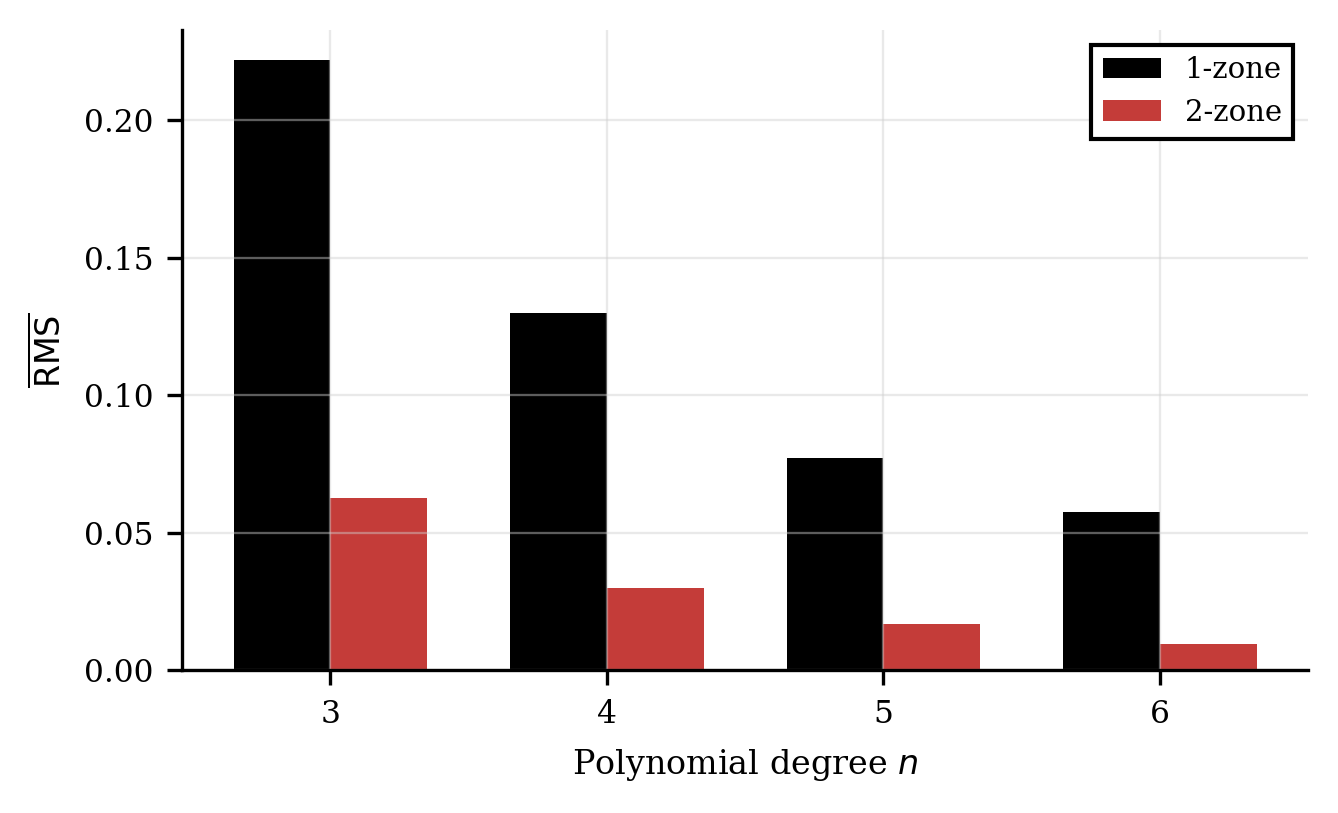}
        \caption{Mean fit error versus polynomial degree $n$ (F4)}
    \end{subfigure}
    \caption{Mean relative RMS error summary for the reduced-order fit study showing error corresponding with a) functional form (F1-F4) and b) polynomial degree order $n$}
    \label{fig:fit_error_summary}
\end{figure}

The results of the brute-force search are shown in Figure \ref{fig:fit_error_summary} and summarized in Table \ref{tab:fit_error_summary}.  In Figure \ref{fig:fit_error_summary}, the functional-form comparison is carried out at fifth order, while the degree scan focuses on the selected reduced-order family, F4.  The error trends show that the 2-zone model significantly outperforms the 1-zone model across the species library.  The transformed functional forms F3 and F4, which fit the natural logarithm of $a_{\mathrm p}$, outperform the direct forms F1 and F2 by a large margin.

The relative increase in average RMS error is fairly small between the fifth- and sixth-order models and the fifth-order model is chosen for parsimony.   The aggregate error is dominated by a small subset of species with sharp spectral-property transitions: SiO has the largest relative RMS (0.0808), followed by AlO (0.0414), TiO (0.0371), and C$_2$ (0.0326), while 20 of the 24 species all have relative RMS errors below 0.03.
The final recommendation is a fifth-order piecewise fit of the form (F4):

\begin{equation}
    a_{\mathrm p}(T)=\exp\left[\sum_{n=0}^{5} c_n \left(\ln(T/1000\,\mathrm{K})\right)^n\right]
\end{equation}

\noindent evaluated separately in two temperature zones with a common breakpoint temperature of $T_{\mathrm{break}}$ = 1500 K and constrained to be continuous at the breakpoint.
Owing to the critical mass of diverse gases, we are confident that future species' absorption curves will be accurately captured using the selected functional form. Table~\ref{tab:final_coefficients} lists the final reduced-order coefficients used for implementation. The piecewise fit is written as two temperature-interval rows per species with the common break temperature at 1500 K. The two species limited to 2500 K, \(\mathrm{C_2H_2}\) and \(\mathrm{CH_4}\), are listed last and retain a single temperature fit.

The final recovery of the mixture-averaged Planck-mean absorption coefficient for use in RTE is thus given by:

\begin{equation}
a_{\mathrm p,\mathrm{mix}}(T) = \sum_{j} p_j \, \exp\left[\sum_{n=0}^{5} c_{n,j} \left(\ln(T/1000\,\mathrm{K})\right)^n\right]
\end{equation}

\noindent where subscript $j$ denotes the $j^{\mathrm{th}}$ species and $p_j$ is its partial pressure in atm, giving $a_{\mathrm p,\mathrm{mix}}$ in units of m$^{-1}$.

\newpage

\begingroup
\scriptsize
\setlength{\tabcolsep}{3pt}
\renewcommand{\arraystretch}{0.9}
\begin{longtable}{@{}llrrrrrr@{}}
\caption{Final pure-gas Planck-mean absorption coefficient curve-fit data: $a_{\mathrm p}(T) = \exp\left[\sum_{n=0}^{5} c_n \left(\ln(T/1000\,\mathrm{K})\right)^n\right]$, with $a_{\mathrm p}$ in units of m$^{-1}$ atm$^{-1}$.}\label{tab:final_coefficients}\\
\toprule
Species & Range [K] & $c_0$ & $c_1$ & $c_2$ & $c_3$ & $c_4$ & $c_5$ \\
\midrule
\endfirsthead

\multicolumn{8}{@{}l}{\tablename\ \thetable\ continued.}\\
\toprule
Species & Range [K] & $c_0$ & $c_1$ & $c_2$ & $c_3$ & $c_4$ & $c_5$ \\
\midrule
\endhead

\bottomrule
\endfoot

\multirow{2}{*}{\(\mathrm{AlF}\)} & 300--1500 & 7.2519458e-01 & -3.3394079e+00 & -4.0704950e-01 & 1.4359252e-01 & -6.6777797e-02 & 1.4709994e-02 \\
 & 1500--5000 & 7.3738880e-01 & -3.4159922e+00 & -2.3366325e-01 & -5.9021988e-03 & -5.7184576e-02 & 2.0087384e-02 \\
\midrule
\multirow{2}{*}{\(\mathrm{AlH}\)} & 300--1500 & 3.1543323e+00 & -2.4061712e+00 & -9.8287733e-01 & 3.8079927e-01 & -1.6769473e-01 & -1.4704280e-03 \\
 & 1500--5000 & -5.9418419e+00 & 6.1471218e+01 & -1.6729605e+02 & 1.9710433e+02 & -1.0360392e+02 & 2.0124764e+01 \\
\midrule
\multirow{2}{*}{\(\mathrm{AlO}\)} & 300--1500 & 3.3185399e+00 & 3.1263786e+00 & -6.2421626e+00 & 2.5847259e+00 & 1.1912063e+01 & 4.6053552e+00 \\
 & 1500--5000 & 3.2802016e+00 & 3.8966633e+00 & -5.0601889e+00 & 1.3536364e-01 & 2.9682830e+00 & -1.1706625e+00 \\
\midrule
\multirow{2}{*}{\(\mathrm{BH}\)} & 300--1500 & 2.7278462e+00 & -1.5533150e+00 & -1.2332456e+00 & 6.4881935e-01 & -1.0072011e+00 & -1.1613523e+00 \\
 & 1500--5000 & -6.4671182e+00 & 6.2874969e+01 & -1.6883316e+02 & 1.9832460e+02 & -1.0407353e+02 & 2.0196056e+01 \\
\midrule
\multirow{2}{*}{\(\mathrm{C_2}\)} & 300--1500 & 3.1953558e+00 & 3.7550092e+00 & -1.7452506e+00 & 1.5192862e+00 & -1.1764626e+01 & -9.6782862e+00 \\
 & 1500--5000 & 6.2253201e-01 & 1.8560632e+01 & -3.5159455e+01 & 3.0771875e+01 & -1.2216448e+01 & 1.7532123e+00 \\
\midrule
\multirow{2}{*}{\(\mathrm{CH}\)} & 300--1500 & 9.0641036e-01 & -9.0608690e-01 & -1.8089830e+00 & 7.8627274e-01 & -5.7749988e-02 & 8.2811814e-02 \\
 & 1500--5000 & -8.3662697e+00 & 6.5247135e+01 & -1.7644361e+02 & 2.0963046e+02 & -1.1098750e+02 & 2.1703720e+01 \\
\midrule
\multirow{2}{*}{\(\mathrm{CN}\)} & 300--1500 & 1.1074649e+00 & 4.7715396e+00 & 6.5841459e+00 & -6.1158142e+00 & -1.6764534e+01 & -7.5913187e+00 \\
 & 1500--5000 & 7.2377651e-01 & 7.8761757e+00 & -3.8322608e+00 & -2.5925640e+00 & 2.6898892e+00 & -5.9491464e-01 \\
\midrule
\multirow{2}{*}{\(\mathrm{CO}\)} & 300--1500 & 8.5962401e-01 & -1.7665210e+00 & -1.3787155e+00 & 5.6532226e-01 & -1.4805834e-01 & 2.0084153e-02 \\
 & 1500--5000 & 8.6160952e-01 & -1.7800784e+00 & -1.3419841e+00 & 5.1886526e-01 & -1.2441147e-01 & 1.3740167e-02 \\
\midrule
\multirow{2}{*}{\(\mathrm{CO_2}\)} & 300--1500 & 3.1667727e+00 & -1.5895193e+00 & -1.4876278e+00 & 6.2820631e-01 & 8.7927462e-02 & -4.9827568e-01 \\
 & 1500--5000 & 3.1965867e+00 & -1.7579362e+00 & -1.0670175e+00 & 7.4573658e-02 & 1.3019515e-01 & -3.4146526e-02 \\
\midrule
\multirow{2}{*}{\(\mathrm{H_2O}\)} & 300--1500 & 1.6444588e+00 & -1.9819290e+00 & -1.6017453e-01 & -1.0847084e-01 & -2.2360369e-01 & -1.7382979e-01 \\
 & 1500--5000 & 1.6136798e+00 & -1.8066382e+00 & -4.7634726e-01 & -9.6593868e-02 & 1.2954393e-01 & -5.0868460e-02 \\
\midrule
\multirow{2}{*}{\(\mathrm{HBO}\)} & 300--1500 & 9.2511824e-01 & -2.2879149e+00 & -8.9777774e-01 & 2.2873010e-01 & 1.6990695e-01 & -3.9361614e-02 \\
 & 1500--5000 & 9.4402100e-01 & -2.2872054e+00 & -1.3002524e+00 & 1.4222046e+00 & -1.1703194e+00 & 2.9321615e-01 \\
\midrule
\multirow{2}{*}{\(\mathrm{HCN}\)} & 300--1500 & 1.1884697e+00 & -1.6402018e+00 & -1.1554654e-01 & -8.5248882e-01 & -2.9650229e-01 & 1.7296054e-01 \\
 & 1500--4000 & 1.0047481e+00 & -3.3077995e-01 & -4.0411770e+00 & 4.6371195e+00 & -3.0210846e+00 & 7.0995695e-01 \\
\midrule
\multirow{2}{*}{\(\mathrm{HF}\)} & 300--1500 & 1.1741417e+00 & -1.0724620e+00 & 4.8316192e-01 & -1.7210634e+00 & -1.4016819e+00 & -2.3344071e-01 \\
 & 1500--5000 & 9.5392329e-01 & 2.8143724e-01 & -3.0575806e+00 & 1.7299597e+00 & -6.8221078e-01 & 1.2024713e-01 \\
\midrule
\multirow{2}{*}{\(\mathrm{MgH}\)} & 300--1500 & 2.1526786e+00 & -2.7342545e+00 & -5.7741210e-01 & 1.8926970e+00 & 1.9156394e+00 & 7.8652640e-01 \\
 & 1500--5000 & 1.1588722e+01 & -6.7943924e+01 & 1.5329337e+02 & -1.4633834e+02 & 6.5647276e+01 & -1.1443040e+01 \\
\midrule
\multirow{2}{*}{\(\mathrm{MgO}\)} & 300--1500 & 5.8408851e+00 & -6.6450087e-02 & -2.4931829e+00 & 7.5636289e-01 & -2.7981595e+00 & -2.5501712e+00 \\
 & 1500--5000 & 3.4982674e+00 & 1.3485443e+01 & -3.1107334e+01 & 2.5684731e+01 & -8.1531154e+00 & 6.4791067e-01 \\
\midrule
\multirow{2}{*}{\(\mathrm{N_2O}\)} & 300--1500 & 2.7456039e+00 & -1.8049970e+00 & -1.2813827e+00 & 4.3752729e-01 & -5.6148526e-02 & -2.0717758e-01 \\
 & 1500--5000 & 3.2272550e+00 & -4.8929045e+00 & 6.1277816e+00 & -7.7669754e+00 & 3.8292386e+00 & -7.3656867e-01 \\
\midrule
\multirow{2}{*}{\(\mathrm{NO}\)} & 300--1500 & 6.3299053e-02 & -2.1012354e+00 & -1.1597885e+00 & 4.8872044e-01 & -1.4886177e-01 & 9.3841655e-03 \\
 & 1500--5000 & 6.5741836e-02 & -2.1198447e+00 & -1.1067288e+00 & 4.0888497e-01 & -8.4908367e-02 & 6.9002025e-03 \\
\midrule
\multirow{2}{*}{\(\mathrm{NO_2}\)} & 300--1500 & 2.5834894e+00 & -2.5476170e+00 & -1.2773552e+00 & -9.2506628e-03 & -4.3408066e-01 & -9.3709822e-02 \\
 & 1500--5000 & 2.5761129e+00 & -2.5457045e+00 & -1.1422229e+00 & -5.3698484e-01 & 2.6572755e-01 & -3.4491369e-02 \\
\midrule
\multirow{2}{*}{\(\mathrm{OH}\)} & 300--1500 & 2.5884282e-01 & -2.4190563e+00 & 2.6573299e-01 & -3.2290045e-02 & 6.4451172e-02 & 9.1216121e-02 \\
 & 1500--5000 & 7.2678718e+00 & -4.9650209e+01 & 1.2095615e+02 & -1.4616042e+02 & 8.2601703e+01 & -1.7265410e+01 \\
\midrule
\multirow{2}{*}{\(\mathrm{SiO}\)} & 300--1500 & 3.7188993e-02 & -2.8322450e+00 & -6.9633147e-01 & -2.2214771e-01 & -9.0645819e-01 & -3.2865030e-01 \\
 & 1500--5000 & 5.4386673e+00 & -4.8368064e+01 & 1.4024076e+02 & -1.9977829e+02 & 1.2929423e+02 & -3.0023804e+01 \\
\midrule
\multirow{2}{*}{\(\mathrm{TiH}\)} & 300--1500 & 2.1793678e+00 & 9.7303477e+00 & -7.7908455e+00 & 2.4751948e+00 & -7.4192075e-01 & 1.3415713e-01 \\
 & 1500--5000 & 9.8022645e-01 & 1.7331509e+01 & -2.5880879e+01 & 2.2576932e+01 & -1.0720619e+01 & 2.0421796e+00 \\
\midrule
\multirow{2}{*}{\(\mathrm{TiO}\)} & 300--1500 & 1.3576596e+00 & -6.6358083e-01 & 8.4240351e+00 & 1.1974477e+01 & 5.5403444e+00 & 5.8694918e-01 \\
 & 1500--5000 & 1.7838808e+00 & -2.0226963e+00 & 2.5102008e+01 & -3.0663070e+01 & 1.5273785e+01 & -2.8738891e+00 \\
\midrule
\(\mathrm{C_2H_2}\) & 300--2500 & 1.6353687e+00 & -2.1687344e+00 & 6.8592879e-03 & -5.5560550e-01 & -1.7116953e-01 & 2.3525066e-01 \\
\midrule
\(\mathrm{CH_4}\) & 300--2500 & 1.1760134e+00 & -1.0994025e+00 & -8.1657403e-01 & -5.2483328e-01 & -2.2529581e-01 & 1.0170867e-01 \\
\end{longtable}
\endgroup

\newpage

\section{Conclusion}\label{conclusion_section}


This work establishes a comprehensive, standardized library of Planck-mean absorption coefficients for pure gases at high temperatures.  By leveraging both the HITEMP and ExoMol databases and solving rigorous line-by-line (LBL) calculations for each species, compact functional forms were generated across a broad temperature envelope of 300 to 5000 K. This dataset extends far beyond standardly available combustion products, equipping modern solvers with the underlying radiative properties of high-temperature radicals, soot precursors, and metal oxides, essential for accurate prediction of radiation in next-generation propulsion, ablation, and hypersonic systems.

Validation was performed against historical data, demonstrating good agreement against legacy fits in the valid temperature ranges, and show high degrees of accuracy when compared against contemporary LBL calculations.
Results show a shift in dominance from stable polyatomic molecules to diatomic radicals and metal-bearing species with increasing temperatures.  Diatomic radicals ($\text{C}_2$, $\text{CN}$) and transition-metal oxides ($\text{TiO}$, $\text{AlO}$) possess strong electronic transition bands in the visible spectrum resulting in Planck-mean absorption coefficients orders of magnitude higher than traditionally considered combustion products (such as CO$_2$ and H$_2$O).
A brute-force search was performed to find an accurate and parsimonious curve fit correlation that covers the entire set of species, enabling the easy assimilation of additional future species. Tabulated coefficients for a fifth-order log-log polynomial curve fit are provided for a total of 24 species.

\appendix
\section*{Appendix}
\section{Numerical and spectroscopic sensitivity}\label{app:sensitivity}

\begin{figure}[htbp]
    \centering
    \includegraphics[width=0.8\textwidth]{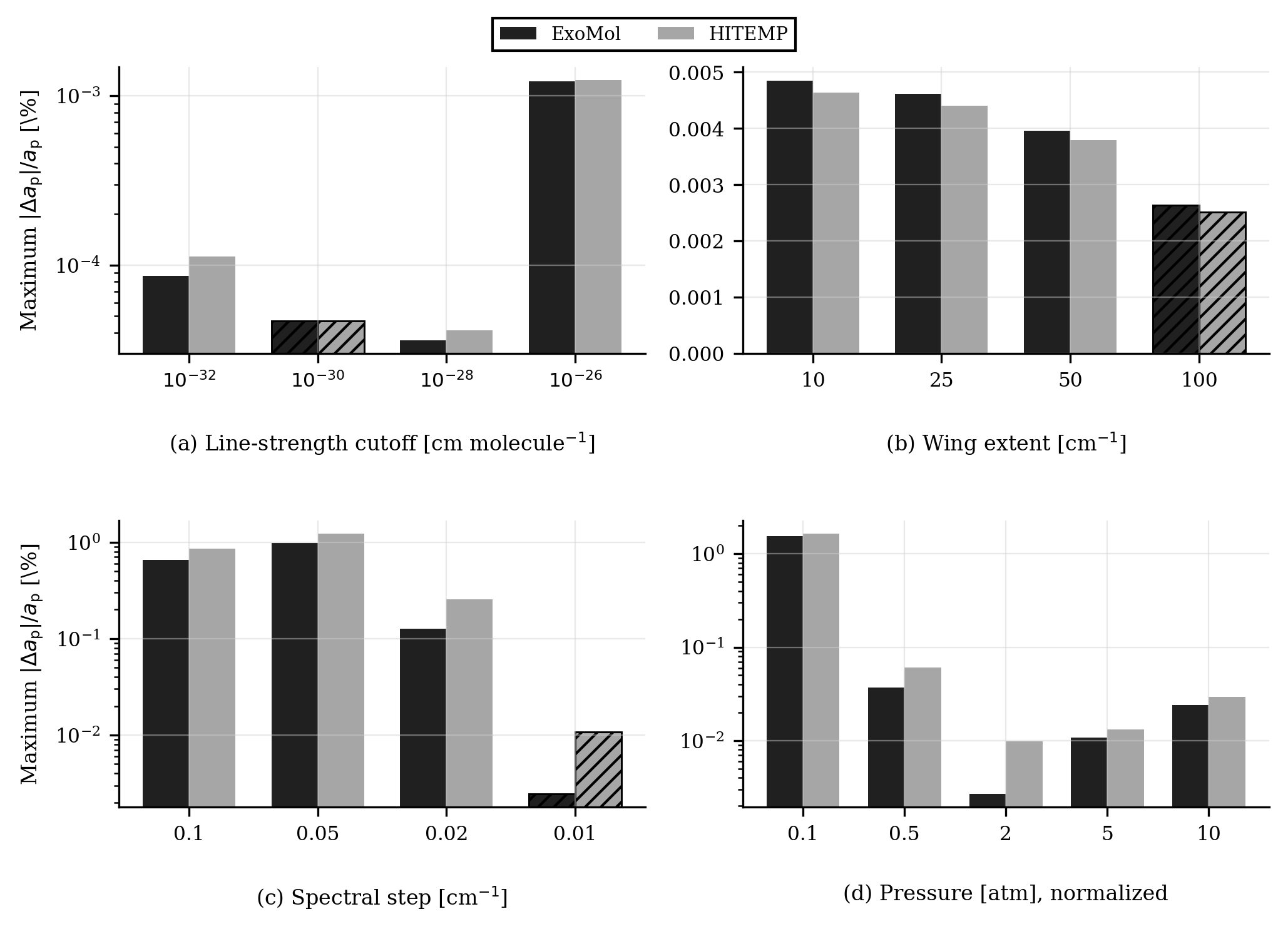}
    \caption{Sensitivity of the CO Planck-mean absorption coefficient to (a) the line-strength cutoff, referenced to no cutoff; (b) the retained line-wing extent, referenced to 200 cm$^{-1}$; (c) the spectral step, referenced to 0.005 cm$^{-1}$; and (d) pressure, after normalization by pressure and referenced to 1 atm.  Hatched lines denote values for final calculations.}
    \label{fig:co_sensitivity}
\end{figure}

The numerical and spectroscopic choices used in the LBL workflow were evaluated using CO as a representative test species. CO was selected because complete high-temperature data was available from both HITEMP and ExoMol and because two ExoMol line-list releases can be compared over the full 300--5000 K temperature interval. Each parameter was varied individually while the remaining production settings were held fixed. Unless it was the parameter under study, the calculation used baseline $S_{\mathrm{cut}}=10^{-30}$ cm molecule$^{-1}$, a 100 cm$^{-1}$ line-wing extent, $\Delta\nu=0.01$ cm$^{-1}$, a Voigt profile, and a total pressure of 1 atm. Figure~\ref{fig:co_sensitivity} reports the maximum absolute relative difference over all temperatures for ExoMol and HITEMP.  Each bar is the maximum absolute relative difference over the entire 300--5000 K interval and the hatched bars identify the numerical settings adopted in the baseline (production) calculations.

The resulting sensitivities demonstrate that the line-strength cutoff and wing extent have negligible effect across the entire tested range.  At the adopted value of $S_{\mathrm{cut}}=10^{-30}$ cm molecule$^{-1}$ and line-wing extent of 100 cm$^{-1}$ results are highly converged.  The adopted spectral step of 0.01 cm$^{-1}$ differs from the 0.005 cm$^{-1}$ reference by at most 0.00246\% for ExoMol and 0.0108\% for HITEMP. Coarser spacing is not uniformly convergent: the maximum differences reach 0.975\% and 1.22\% at 0.05 cm$^{-1}$ for ExoMol and HITEMP, respectively. These results support the selected cutoff, wing extent, and spectral resolution without incurring the cost of the most conservative reference settings.  Pressure sensitivity was evaluated after dividing $a_{\mathrm p}$ by pressure to remove the leading ideal-gas number-density dependence.  Across all pressures, the maximum differences are less than 1.6\%, with artificial errors present at the lowest pressure due to the difficulty of resolving narrower profiles on the fixed spectral grid.  The results support linear extrapolation within low to moderate pressures.

The sensitivity was highest across spectroscopic sources.  Within ExoMol, the `Li2015' ground-state rovibrational list \cite{Li2015RovibrationalLine} produces values 1.369--1.409\% above the most-current `4thplus' list \cite{tao_CO_exomol}, with a mean difference of 1.382\%.  This largely accounts for the minor variation observed in historical fits using legacy lists and procedural methods.
Notably, under the same baseline settings and line list \cite{Li2015RovibrationalLine}, calculations performed using the ExoMol and HITEMP databases predicted maximum differences less than 0.005\% owing to minor discrepancies in collisional broadening and pressure-shift treatment.

\section*{Acknowledgements}

This work was sponsored by the Office of Naval Research through the NRL 6.1 Base Program and the Hypersonic Aerothermodynamics, High-Speed Propulsion and Materials portfolio (PM: Dr. Eric Marineau). R.D.D. was supported through the National Research Council postdoctoral associateship awards administered by the National Academies of Sciences, Engineering, and Medicine.  The authors would like to thank Mr. Keith Obenschain for help with managing computational resources.

\section*{Declaration of Competing Interests}

The authors declare that they have no known competing financial interests or personal relationships that could have appeared to influence the work reported in this paper.

\bibliographystyle{unsrtnat}
\bibliography{citations}

\end{document}